\documentclass[sigconf]{acmart}

\usepackage{algorithmic}
\usepackage{graphicx}
\usepackage{textcomp}
\usepackage{xcolor}
\usepackage{booktabs} 
\usepackage{soul}
\graphicspath{{figs/}}
\usepackage{braket}
\usepackage{etoolbox}

\copyrightyear{2022} 
\acmYear{2022} 
\setcopyright{acmcopyright}\acmConference[GLSVLSI '22]{Proceedings of the Great Lakes Symposium on VLSI 2022}{June 6--8, 2022}{Irvine, CA, USA}
\acmBooktitle{Proceedings of the Great Lakes Symposium on VLSI 2022 (GLSVLSI '22), June 6--8, 2022, Irvine, CA, USA}
\acmPrice{15.00}
\acmDOI{10.1145/3526241.3530833}
\acmISBN{978-1-4503-9322-5/22/06}
\AtBeginDocument{%
  \providecommand\BibTeX{{%
    \normalfont B\kern-0.5em{\scshape i\kern-0.25em b}\kern-0.8em\TeX}}}

\begin{document}
\fancyhead{}

\title[Security Aspects of Quantum Machine Learning]{Security Aspects of Quantum Machine Learning: Opportunities, Threats and Defenses (Invited)}


\author{Satwik Kundu}
\affiliation{%
  \institution{The Pennsylvania State University}
  \city{State College}
  \state{Pennsylvania}
  \country{USA}
  \postcode{16801}}
\email{sxk6259@psu.edu}

\author{Swaroop Ghosh}
\affiliation{%
  \institution{The Pennsylvania State University}
  \city{State College}
  \state{Pennsylvania}
  \country{USA}
  \postcode{16801}}
\email{szg212@psu.edu}


\begin{abstract}
  In the last few years, quantum computing has experienced a growth spurt. One exciting avenue of quantum computing is quantum machine learning (QML) which can exploit the high dimensional Hilbert space to learn richer representations from limited data and thus can efficiently solve complex learning tasks. Despite the increased interest in QML, there have not been many studies that discuss the security aspects of QML. In this work, we explored the possible future applications of QML in the hardware security domain. We also expose the security vulnerabilities of QML and emerging attack models, and corresponding countermeasures.
\end{abstract}



\maketitle

\section{Introduction}
Quantum computing is a new computing paradigm with enormous potential for the future. Despite the fact that the technology is still in its infancy, the community is looking for computational advantage from quantum computers (i.e., quantum supremacy) for practical applications such as material/drug discovery \cite{li2021drug} \cite{li2021quantum}. In the near future, Quantum Machine Learning (QML) is a promising application domain for achieving quantum advantage with noisy quantum computers. Image classification could be transformed by QML. Several Parameterized Quantum Circuit (PQC) based QML models, also known as Quantum Neural Network (QNN), have already been proposed in the literatures \cite{schuld2021effect}. A traditional QNN consists of a data encoding circuit, a PQC, and measurement operations that can be trained to perform traditional Machine Learning (ML) tasks such as classification, regression, distribution generation, and so on.

However, the security of QML/QNNs against adversaries, as well as their effectiveness in solving security problems remains unexplored. This may come as a surprise given the long history of quantum computing applications in security. With more and more ML algorithms being used by system defenders and attackers to secure and attack hardware, we attempt to investigate how QML can be used to assist in the resolution of hardware security issues. This is primarily due to the fact that hardware supply chain is plagued by threats like counterfeiting, Trojan insertion and tampering. A slight error in ML based detection approaches may lead to compromised hardware in mission critical systems. Hence, there is a need to explore a fundamentally superior ML model such as, QML.
Although attractive, quantum classifiers, like classical neural network based classifiers, are also vulnerable to carefully crafted adversarial examples, which are obtained by adding imperceptible perturbations to legitimate input data (adversarial input manipulation). 

In this paper, we investigate the use of QML with a hybrid quantum-classical model, specifically classical dimension reduction + QNN, to classify Printed Circuit Board (PCB) defects, which have become a pressing need in the PCB industry because they can severely affect system performance/security. We also discuss how this model can be utilized for hardware Trojan and recycled chip detection, which makes the electronic supply chain untrustworthy. We also study potential vulnerabilities and threats to QML models, as well as numerous approaches to minimize them.

The rest of the paper is structured as follows; We cover basics on quantum computing and QNN in Section \ref{preli}, discuss the security applications of QML in Section \ref{secappli}, present various vulnerabilities and attack models on QML in Section \ref{attacks}, and draw the conclusions in Section \ref{conc}.

\begin{figure*}[!t]
    \vspace{-4mm}
     \begin{center}
        \includegraphics[width=0.9\textwidth]{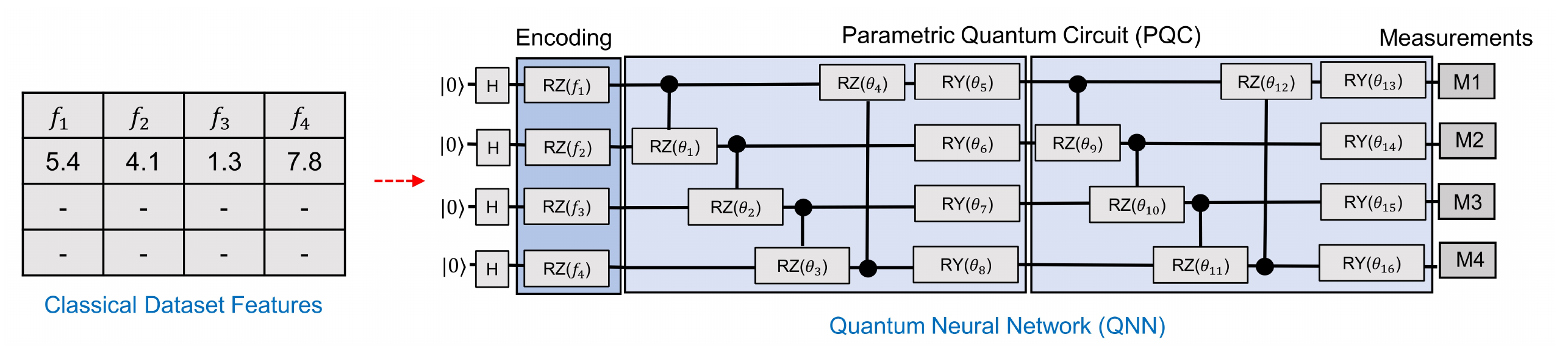}
     \end{center}
     \vspace{-4mm}
    \caption{Example hybrid quantum-classical architecture. Classical features are transformed to quantum states using angle encoding where each feature is encoded into single qubit (1:1 encoding for example, RZ($f_1$)). The encoded quantum states are then subjected to multiple transformations by a parameterized circuit before being measured.
    } 
    \label{fig:arch}
    \vspace{-2mm}
\end{figure*}

\section{Quantum Computing Basics} \label{preli}

\subsection{Qubits, Quantum Gates \& Measurements} A qubit is a fundamental building block of a quantum processor and are usually driven by microwave pulses. It is a two-level system that stores data as quantum states. A qubit is analogous to a classical bit. However, unlike a classical bit, a qubit can be in a superposition state, which is a combination of $\ket{0}$ and $\ket{1}$ at the same time. Mathematically, the qubit state is represented by state vector $\ket{\psi} = a\ket{0} + b\ket{1}$ where $|a|^2$ and $|b|^2$ represent probabilites of `0' and `1' respectively (thus, $|a|^2 + |b|^2 = 1$). A variety of qubit technologies exists, e.g., superconducting qubits, trapped-ions, neutral atoms, silicon spin qubits, to name a few \cite{nielsen2002quantum}. 

Quantum gates are operations that change the state of qubits, allowing them to perform computations. Quantum gates can operate on a single qubit (for example, the X (NOT) gate) or on multiple qubits (e.g., 2-qubit CNOT gates). They are realized physically through the use of pulses (e.g., laser pulse in Ion Trap qubits, RF pulse in Superconducting qubit, etc.) A quantum circuit can have a large number of gate operations. Qubits are measured in a desired basis to determine the final state of a quantum program. Measurements in physical quantum computers are typically restricted to a computational basis, such as the Z-basis in IBM quantum computers.

\textit{Measurement/Expectation Value} is another important concept in quantum computing. It is the average of the eigenvalues, weighted by the probabilities that the measured state is in the corresponding eigenstate. Mathematically, expectation value of an operator ($\sigma$) is defined as $\langle\psi|\sigma|\psi\rangle$ where $|\psi\rangle$ is the qubit state vector. It varies between the minimum and maximum eigenvalues of the operator. For example, the Pauli-Z ($\sigma_z$) operator has two eigenvalues: +1 and -1. Therefore, the Pauli-Z expectation value of a qubit will vary in the range of [-1, 1] depending on the qubit state.

\subsection{Quantum Noise} In theory, quantum computers should provide exponential speedups over classical computers in a variety of tasks; however, this has not been the case in practice. Among many impeding factors, quantum computers suffer from a wide range of error/noise modes, such as, a qubit can only retain its state for a short period of time which leads to Coherence errors. Quantum gates are realized using microwave/laser pulses and thus it is impossible to precisely generate and apply these pulses in actual hardware which is known as Gate errors. Measurement errors occurs when a $\ket{0}$ state qubit is measured as $\ket{1}$ due to imprecise measurement apparatus and, Parallel execution of multiple gates on different qubits can affect each other's performance which is known as Crosstalk errors. The rates of these errors vary depending on the qubits and hardware used, and can impede the performance of QML models' classifiers, resulting in unreliable outputs and additional security risks.

\subsection{Quantum Neural Networks (QNN)} QNN entails optimizing the parameters of a Parametric Quantum Circuit (PQC) to achieve the desired input-output relationship. QNN is typically divided into three sections: (i) a classical to quantum data encoding (or embedding) circuit, (ii) a parameterized quantum circuit, and (iii) measurement operations. A variety of encoding methods are available in the literature \cite{schuld2021effect}. For continuous variables, the most widely used encoding scheme is angle encoding where a continuous variable input classical feature is encoded as a rotation of a qubit along the desired axis \cite{abbas2020power, schuld2020circuit, schuld2021effect}. Thus, to encode `n' classical features, we require `n' qubits. For example, RZ($f_1$) on a qubit in superposition (Hadamard - H gate is used to put the qubit in superposition) is used to encode a classical feature `$f_1$' in Fig. \ref{fig:arch}. We can also encode multiple continuous variables in a single qubit using sequential rotations. As the states produced by a qubit rotation along any axis will repeat in 2$\pi$ intervals, features are generally scaled within 0 to 2$\pi$ (or -$\pi$ to $\pi$) in a data pre-processing step.

{\bf{Parametric Quantum Circuit (PQC):}} The parametric circuit is composed of two components: entangling operations and parameterized single-qubit rotations. The entanglement operations are a set of multi-qubit operations performed between all of the qubits to generate correlated states \cite{lloyd2020quantum}. To search the solution space, the parametric single-qubit operations listed below are used. In QNN, this combination of entangling and single-qubit rotation operations is referred to as a parametric layer. It is worth noting that substantial research has been conducted in recent years to determine the best PQC architecture for QNN.
Descriptors such as expressive power, entanglement capability, effective dimension, and so on have been proposed to assess the efficacy of various PQC options \cite{sim2019expressibility, abbas2020power}. Proponents of these descriptors assert that there is a significant relationship between the descriptor values and the trainability of quantum circuits. Such descriptors may be useful in practical applications to select the best PQC architecture for the intended QML application.

{\bf{Cost Functions:}} The output state of a QNN circuit is determined by measuring qubits in the computational basis. The network is trained using a cost function derived from the measurements \cite{abbas2020power, schuld2020circuit}. In a binary classification problem, for example, the authors measured all the qubits in the QNN model in Pauli-Z basis and associated class 0 with the probability of obtaining even parity and class 1 with the probability of obtaining odd parity \cite{abbas2020power}. Then, the model is trained using binary cross-entropy loss. In \cite{alam2019addressing}, the authors used the Pauli-Z expectation value of a single qubit (-1 associated with class 1 and +1 associated with class 0) for a binary classifier and trained it using mean squared error (MSE) loss.
In \cite{huang2021power}, the authors fed the outputs of the QNN to a classical neural network and trained it using the binary cross-entropy loss function.

{\bf{Training:}} QNNs can be trained using any gradient-based optimization algorithm such as, Adam \cite{kingma2014adam} or Adagrad \cite{duchi2011adaptive}. To apply these methods, we need to compute the gradients \cite{banchi2021measuring, schuld2019evaluating} of the QNN outputs with respect to the circuit parameters. The parameter-shift rule is a known method to compute the gradients \cite{banchi2021measuring, schuld2019evaluating}. The parameter-shift rule is conceptually very similar to the age-old finite difference method, which uses two close-proximity evaluations of a target function to compute gradients with respect to a parameter. In the parameter-shift rule, the two data points can be far apart, unlike in the finite difference rule. As a result, it is more resistant to shot noise and measurement errors than finite difference. \cite{schuld2019evaluating}. Alternatively, one can also use gradient free optimizer such as Nelder-Mead to train a QNN \cite{lavrijsen2020classical}. A gradient-free optimizer, on the other hand, may perform poorly when the network has a large number of parameters.

\section{Security Applications} \label{secappli}

In this section, we explore the usage of QML in defect classification of PCB and discuss other possible security applications.

\subsection{PCB Defect Classification}
{\bf{Overview:}}
PCBs are the fundamental building blocks of the majority of modern electronic devices. However, due to limited technology, a 100\% quality rate cannot be guaranteed in the PCB production process. Sometimes the PCBs will have short circuit, missing hole, spur, and other defects. Because the task of manufacturing a large number of PCBs is frequently outsourced to third-party vendors in order to save money, these PCBs are vulnerable to attacks. For example, defects could be purposefully introduced by these untrustworthy vendors, rendering PCBs faulty and severely disrupting the workflow of systems that rely on these PCBs.
We attempt to demonstrate a potential security application of QML in classifying various PCB defects using a hybrid quantum-classical model. For this task, we employ the framework proposed by \cite{mahabubulqnn}, in which we first use a convolutional autoencoder to reduce dimensionality before training a QNN with the crucial extracted features for our classification problem.

{\bf{Convolutional Autoencoders (CAE):}} Autoencoders (AE) are a type of feedforward neural network. They use an encoder network to compress the input into a lower-dimensional code, and then use a decoder network to reconstruct the output from this representation. To train the network, the distance between the input and the reconstructed output (e.g., MSE loss) is used as the feedback signal. However, CAE has a better architecture than AE for extracting textural features from images. In CAE, the encoder block begins with one or more successive convolutional layers. The decoder block concludes with convolutional transpose/deconvolutional layers. In the center is a fully connected AE, the innermost layer of which is made up of a small number of neurons. Once trained, the encoder block can be used as a stand-alone entity to extract a lower-dimensional representation of the input data.

\begin{figure}[!t]
     \begin{center}
        \includegraphics[width=0.4\textwidth]{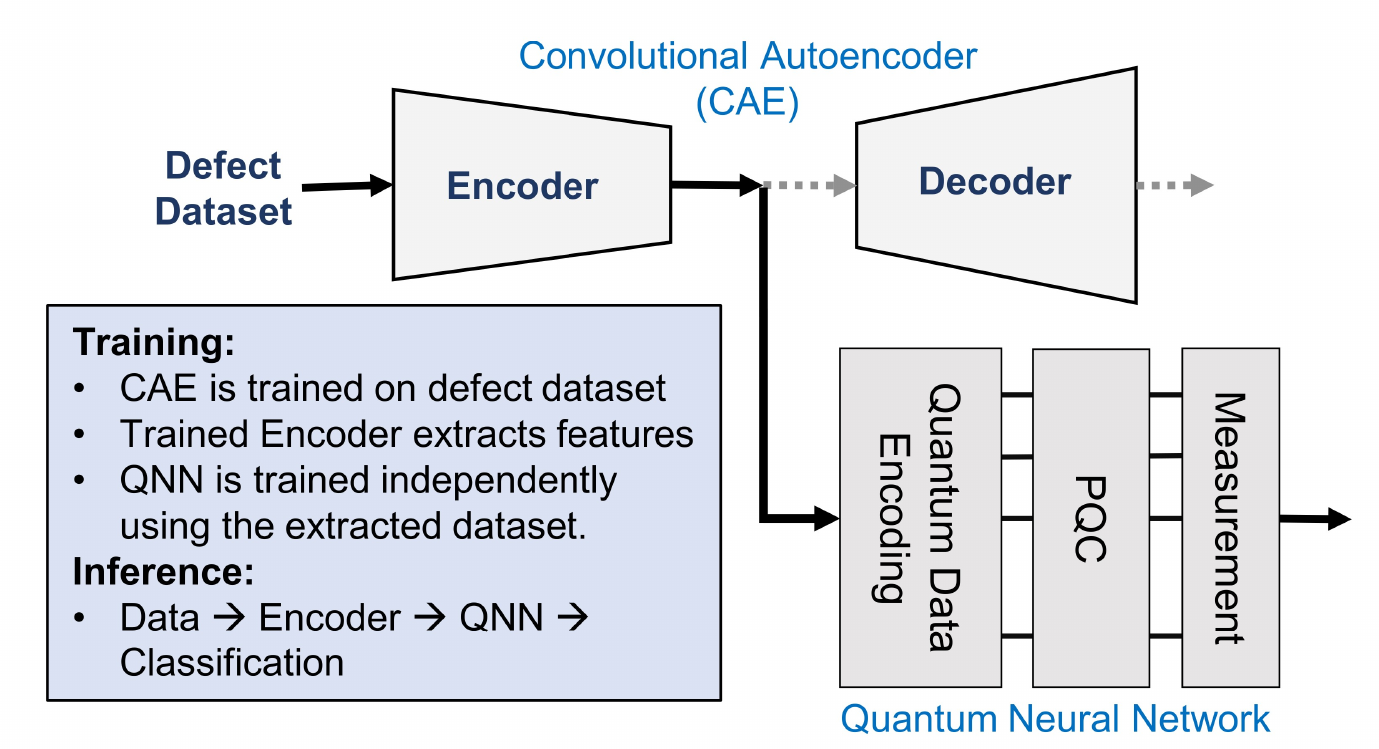}
     \end{center}
     \vspace{-4mm}
    \caption{The network architecture of CAE + QNN. The trained CAE Encoder block generates a lower-dimensional image representation for the QNN.} 
    \label{fig:cae}
    \vspace{-2mm}
\end{figure}

Fig. \ref{fig:cae} shows how CAE is used in this work (for PCB defect classification). The final ConvTranspose2d layer uses Sigmoid activation where `$d$' is the dimension of the latent-space.

{\bf{Classification Model:}} The hybrid network (Fig. \ref{fig:cae}) consists of two separate networks: a CAE and a QNN. To learn a lower dimensional representation of the data, the CAE is trained with the original image dataset. To extract image features, the trained encoder network is used. To perform final classification, a conventional QNN is trained with these extracted features and image labels.
This architecture is named as CAE+QNN.

In terms of encoding methods, parametric circuits, and measurement operations, a QNN/quantum filter has a plethora of design options. The Python framework developed by \cite{mahabubulqnn} supports a wide range of these options, which will impact the learnability of the QNN \cite{sim2019expressibility}. However, we use the single feature/qubit encoding method like in (Fig. \ref{fig:arch}), the PQCircuit-16 from \cite{hubregtsen2021evaluation}, and Z-basis measurements of the qubits in the QNN. In addition, we limit the number of trainable parametric layers to two. We feed the QNN outputs to a fully-connected layer \cite{huang2021power}. The number of output neurons is equal to the number of classes.

\begin{table}[!b]
    \vspace{-2mm}
  \caption{Augmented PCB Defect Dataset}
  \vspace{-2mm}
  \label{tab:freq}
  \begin{tabular}{cc}
    \toprule
    Defects & \# of images\\
    \midrule
    Missing hole & 3612\\ 
    Mouse bite & 3684 \\
    Open circuit & 3548 \\
    Short & 3508 \\
    Spur & 3636 \\
    Spurious copper &3676 \\
  \bottomrule
\end{tabular}
\label{table:1}
\end{table}

\begin{figure}[!t]
     \begin{center}
        \includegraphics[width=0.4\textwidth]{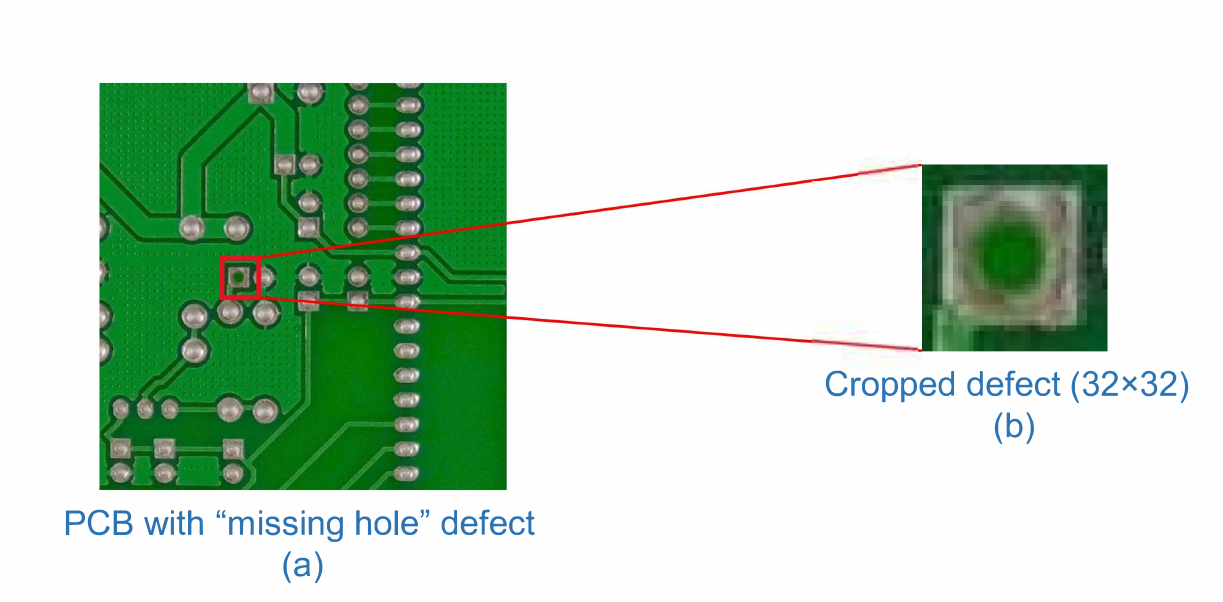}
     \end{center}
     \vspace{-2mm}
    \caption{(a) Original PCB image of size $600\times600$ (b) Cropped defect image of size $32\times32$ used for classification}
    \label{fig:defect}
    \vspace{-8mm}
\end{figure}

{\bf{Dataset:}} We chose the augmented PCB defect dataset, which was originally released by Peking University's Open Lab on Human Robot Interaction, and later \cite{tddnet} performed augmentation techniques on the original dataset to create this new dataset. It contains 6 types of defects which are created by photoshop using Adobe softwares. The defects defined in the dataset are: missing hole, mouse bite, open circuit, short, spur, and spurious copper. The dataset contains a total of 10,668 images of size $600\times600$, each of which may or may not have more than one defect, but all defects in an image belong to the same class. For our PCB defect classification task, we further cropped out these defects from each of these images to form a larger dataset of 21,664 images, each of size $32\times32$ (Fig. \ref{fig:defect}). More information regarding no. of images per defect is shown in Table \ref{table:1}. We first trained our convolutional autoencoder with our defect dataset images (Train:Test = 70:30). Later, we created one smaller 3-Class and one 6-Class classification datasets, using the trained model with latent dimension(d) = 4 which we used to train our QNN (for classification).

{\bf{Training Setup/Metrics:}} We trained our CAE with 15,165 training samples from the PCB defect dataset with latent dimension $d = 4$ [\textit{Loss function:} Mean Squared Error (MSE), \textit{Optimizer:} Adam, \textit{Learning rate:} 0.001, \textit{Epochs:} 25, and \textit{Batch size:} 50]. We tested this trained CAE using the testing samples to generate the reduced feature set (of $d = 4$). This reduced dataset is further divided into two equal sets (3 \& 6 class) of 2000 samples each. Finally, these sets are used for QNN training and validation (70:30 split).

To evaluate the performance of the QNN model, we use the average accuracy over the entire training and validation dataset \cite{anthony2009neural}. The training accuracy of the models indicates their trainability, whereas the validation accuracy indicates their generalization capability. To train our model, we use the gradient-based Adagrad optimizer \cite{paszke2019pytorch}. We used angle encoding technique to encode the classical features, encoding one feature per qubit, as shown in Fig. \ref{fig:arch}. Thus, for our four feature dataset, we used four qubits. In our 4-qubit model, we limited the number of parametric layers to two. We trained the QNN for 10 epochs [\textit{Loss function:} Sparse Categorical Cross Entropy, \textit{Optimizer:} Adagrad, \textit{Learning rate:} 0.5, and \textit{Batch size:} 32].

{\bf{Results:}} The performance after 10 epochs of training are tabulated in Table \ref{table:2}. The chosen number of latent-dimension ($d$) dictates the QNN architecture. It also affects the overall network performance. A higher value of $d$ means more input features for the QNN model that generally translates to better training performance of the QNN. Therefore, a higher $d$ (at the cost of larger QNN) may provide better performance in practical applications. Also, there are myriad of PQC options to choose from hence there might be some other PQC which provides better classification rates; the analysis by authors in \cite{hubregtsen2021evaluation} shows the PQCircuit-5 \& 6 with the high number of gate count (trainable parameters) consistently provides higher accuracy at cost of higher training time, when compared to the other PQCs.

\begin{table}[!b]
    \vspace{-6mm}
    \centering
    \caption{CAE+QNN Architecture performance after 10 Epochs}
    \vspace{-2mm}
    \begin{tabular}{c|c|c} 
        \textbf{Dataset} & \textbf{Train Accuracy} & \textbf{Val. Accuracy}\\
        \hline
        Defect 3-Class & 0.70 & 0.68 \\
        Defect 6-Class & 0.46 & 0.42 \\
    \end{tabular}
    \label{table:2}
\end{table}

It should be noted that the goal of this work is not to demonstrate superior classification accuracy over our classical counterparts, but rather to demonstrate the potential applications of QML.

\subsection{Hardware Trojan \& Recycled Chip Detection}
 
Hardware Trojan attacks have emerged as a significant security risk for Integrated Circuits (IC). In simple terms, a Hardware Trojan attack is a malicious intentional modification of an electronic circuit or design that results in unfavorable behavior and leads to security threats. Over the years, several classical machine learning techniques have been explored across different abstraction levels to detect this malicious hardware virus such as; (a) Reverse Engineering\cite{reverseNasr}, (b) Real time detection\cite{kulkarni}, (c) Gate-Level Netlists Detection \cite{hasegawa} etc.
Another important concern for the security and dependability of electronic systems/devices is the use of recycled ICs. Using counterfeited/recycled ICs might pose a major threat since during the recycling process, an adversary could add a Trojan, change the functionality, or even insert faults, which could lead to important information leakage or just impede the device's performance. 

The proposed hybrid quantum-classical framework (or any another QML model) can be trained with proper datasets and data preparations/pre-processing to detect the Trojans and recycled parts.

\subsection{Usage Model of QML in Hardware Security}
QML can be used in hardware security domain in two ways, (i) full quantum approach, where the QML is used to screen all PCBs, recycled chips and Trojan infested designs. This approach will be costly since quantum computers are expensive but will provide best security guarantees, (ii) hybrid approach, where classical ML is used to screen the PCBs, chips and designs first and only the chips dedicated for mission critical systems are forwarded to the QML.  

\section{Quantum Machine Learning: Vulnerabilities, Attacks \& Defenses}\label{attacks}
By utilizing quantum mechanics principles such as superposition, tunneling, and entanglement QML have given hope of outperforming their classical counterparts in the near term, even with noisy intermediate-scale quantum (NISQ) hardware. The high-dimensional Hilbert space of sizable quantum systems provides a naturally advantageous starting ground for QML models for classification tasks where statistical patterns can be revealed in complex feature spaces. However, QML models, like many state-of-the-art classical machine learning models, possess assets and are vulnerable to attack. Researchers have demonstrated that the adverserial robustness of any classifier is increasingly reduced by the dimensions of the space on which it classifies. This has caught the interest of QC/QML researchers, as QML models take advantage of quantum systems' high dimensionality. In this section, we discuss the vulnerabilities, possible attacks that could compromise QML's Intellectual Property (IP), reliability and performance.

\subsection{Assets}
QML circuits possess following assets, (i) training data embedded in state preparation circuit; (ii) type of encoding used in the state preparation circuit; (iii) PQC ansatz; (iv) number of parameters; (v) number of qubits and number of PQC layers.

\subsection{Vulnerabilities}

{\bf{Identical coupling hardware in quantum cloud:}} Several companies, including IonQ, D-Wave, IBM, and Rigetti, now provide access to their quantum computers via cloud-based partners such as Azure Quantum, Amazon Braket, Google Cloud etc, where researchers frequently deploy their QML models to evaluate the robustness/performance in noisy environments. These cloud service providers frequently have multiple hardware with varying hardware quality and architectures, such as the number of qubits, coupling map, and so on. The scheduler at the cloud service provider's end may have multiple hardware with the same coupling map at times. Furthermore, coupling maps of larger hardware, such as ibmq\_rochester, with a greater number of qubits, can fit the coupling map of many smaller hardware, such as ibmq\_london and ibmq\_santiago. Therefore, many choices of the user defined coupling map architecture exists in the quantum cloud. Unfortunately, user lacks capability to distinguish the identical coupling maps at cloud end. As a result, third-party cloud vendors may assign low-quality hardware to the job, resulting in poor results and/or a longer convergence time for the quantum circuit.

{\bf{Reliance on compilation quality:}} Quantum circuit compilation is another important step in converting quantum programs written in high-level gate sets to low-level (native) gate sets tailored to the underlying hardware. The compilation is also crucial to the program's success on real-world hardware. Several companies nowadays have their own quantum hardware frequently offer compilers for their hardware, such as D-Wave's Ocean, IBM's Qiskit, Rigetti's QuilC compilers etc, as well as some third-party software tools, including compilers, such as Orquestra and tKet. With the growing interest in quantum computing, in future there may be several third-party compilers that provide better performance due to better optimization algorithms, etc. However, compilers from these less trustworthy companies would eventually lead to security and privacy issues.

{\bf{Embedding of sensitive private property in QML circuit:}} 
QNN circuits with novel PQC or angle encoding techniques are frequently designed and tested on real quantum hardware. In order to take advantage of increased speed/efficiency, they might be sent to third-party compilers, who would have complete access to the circuit architecture, posing a threat to intellectual property. Likewise, the quantum circuit may also include sensitive information such as financial analysis and proprietary algorithms thus it might not be the best strategy to directly send the raw quantum circuit for testing to/through untrusted compilers/cloud providers.

\subsection{Attack Models and Defenses}

{\bf{Unreliable Hardware Allocator:}} As previously stated, untrustworthy quantum computers from third parties can allocate poor quality hardware to save money or to meet their falsely advertised qubit or quantum hardware specifications, severely degrading QML model performance.

In \cite{qpuf}, authors proposed Quantum Physically Unclonable Function (QuPUF) to verify the identity of the hardware assigned by the cloud-based scheduler before sending the actual workload. As each qubit is distinct in term of gate error, readout error, decoherence error, their idea is to design a QuPUF to convert these error rates into qubit signature, which forms the hardware signature. The work used various types of QuPUFs like; (a) \textit{Hadamard gate-based QuPUF} where qubits are placed in a superposition state using $\mathbb{H}$-gate followed by the measurement. Qubit values are expected to be biased towards either zero or one, depending on the errors that act as unique device signature, (b) \textit{Decoherence-based QuPUF} which uses the difference in the decoherence times of the qubits to generate a response etc. Experiments on real IBM quantum hardware revealed that QuPUF achieves an inter-die Hamming Distance (HD) of 55\% and an intra-HD of 4\%, compared to ideal cases of 5nd 0\%, respectively. QuPUF can also be used to defend against unreliable hardware allocator attack for QML application.

{\bf{Compilation Oriented Attacks:}} We also discussed earlier how sending a novel QML architecture or algorithm to untrusted compilers creates opportunities for adversaries to steal IP namely, the type of ansatz used, the number of parameters, the number of layers in the QNN, the type of input embedding, to name a few. The adversary can also insert malicious components in the QNN to poison training and lead to misclassification. 

The work in \cite{saki-split}, presents a novel split methodology to secure IPs from untrusted compilers while taking advantage of their optimizations. The main idea behind their proposed method is that instead of sending the entire quantum circuit at once, it is divided into multiple parts that are sent to a single compiler at different times or to multiple compilers, and these sub-circuits are later combined together accordingly by the designer post-compilation. They conducted extensive experiments with 152 circuits and concluded that this split compilation method can completely secure IPs or introduce factorial time reconstruction complexity with a minor overhead (max 6\%). 

In \cite{aakarshitha}, to make circuits more robust against any kind of tampering/counterfeiting from untrusted third party compilers, they insert dummy SWAP gates to corrupt the functionality of the program. Unlike classical chips, quantum chips do not reveal circuit functionality, so the adversary cannot estimate the SWAP gate location since, any gate can be a potential dummy gate making it hard for the adversary to reverse engineer the quantum circuit. They presented a method for determining the best position for dummy SWAP gate insertion that maximizes Total Variation Distance (TVD) without requiring time-consuming quantum circuit simulation. Experiments show that their proposed metric achieved a $\approx 6\% $ improvement over average TVD and a $\approx 12\% $ improvement over best TVD with minimal overhead. As a result, third-party cloud suppliers may assign low-quality hardware to the job. 

In case of QNNs, we may split the state preparation circuit and PQC for compilation in separate vendors. One may also split the PQC further for compilation in randomized order. This approach will protect the information about state preparation circuit and PQC. For further obfuscation (or as an alternate standalone defense technique), dummy ZZ gates can be inserted in the PQC layer to alter the ansatz and number of parameters in QNNs architecture. Even though the adversary will obtain the QNN, the training and performance of the obfuscated QNN will be different than the unobfuscated QNN. A careful designer can even place the dummy ZZ gates to cause barren plataue in the solution space to make the QNN useless.

{\bf{Fault Injection Attacks:}} The gate error of an isolated operation may differ from the gate error with another gate operation in parallel. This is known as the crosstalk error. In \cite{ASPLOS-2020-MuraliMMJ}, researchers showed that the gate error with another operation in parallel can be $\approx 3$ times higher than an isolated gate operation. As a result, crosstalk may negatively impact the prediciton/classification accuracy of a QML model. Furthermore, it has recently been demonstrated that an external adversary can inject faults into another user's program sharing the same hardware by repeatedly driving qubits using CNOT gates \cite{saki2020noise}. This would cause a QNN to perform worse than it would in a normal environment.

Saki \textit{et al.} \cite{saki2020noise}, investigated these types of threats on quantum circuits and classifiers, and proposed a method to mitigate the aforementioned threat by introducing isolation/buffer qubits between user programs. This basically means that when one QNN circuit is running on a set of qubits in a hardware, another program is not allowed to run on the previous program's neighbouring qubits, which is maintained by inserting/considering those neighbouring qubits as buffer/isolation qubits. Their analysis shows that buffer qubits provide as much as 1.87$\times$ higher reliability at the cost of a few unused qubits.

\section{Conclusion} \label{conc}

Machine learning techniques are widely used in almost every aspect of modern society, making it a popular area of scientific research in terms of both attack and defense mechanisms. With the recent surge in research interest in Quantum Machine Learning (QML), we investigate the security opportunities, vulnerabilities, threats, and defenses of QML. We present a potential security application of QML in detecting PCB defects and suggest hardware Trojan and recycling chip detection as additional application areas to secure the semiconductor supply chain. We also describe the QML's vulnerabilities, attack models and defenses.

\begin{acks}
The work is supported in parts by NSF (CNS-1722557, CCF-1718474, OIA-2040667, DGE-1723687 and DGE-1821766) and seed grants from Penn State ICDS and Huck Institute of the Life Sciences. Also, a special thanks to Dr. Md Mahabubul Alam for his guidance and for providing the CAE+QNN codebase.
\end{acks}

\bibliographystyle{ACM-Reference-Format}
\bibliography{acmart}

\end{document}